\documentclass[preprint]{aastex62}
\usepackage{xcolor}
\usepackage{multirow}
\usepackage[normalem]{ulem}
\submitjournal{ApJ}
\accepted{\today}

\begin{document}

\title{The Trigger Mechanism of Recurrent Solar Active Region Jets Revealed by the Magnetic Properties of a Coronal Geyser Site}

\correspondingauthor{Alin Razvan Paraschiv}
\email{alin.paraschiv@monash.edu; paraschiv.alinrazvan@gmail.com}

\author[0000-0002-3491-1983]{Alin Razvan Paraschiv}
\affiliation{School of Mathematical Sciences, Monash University \\
9 Rainforest Walk, Clayton, Victoria 3800, Australia}

\author[0000-0002-4111-3496]{Alina Donea}
\affiliation{School of Mathematical Sciences, Monash University \\
9 Rainforest Walk, Clayton, Victoria 3800, Australia}

\author[0000-0003-0026-931X]{K.D. Leka}
\affiliation{NorthWest Research Associates\\
3380 Mitchell Lane, Boulder, Colorado, USA}
\affiliation{Nagoya University / ISEE\\
Furo-cho Chikusa-ku, Nagoya, Aichi 464-9601 JAPAN}

\begin{abstract}
Solar active region jets are small-scale collimated plasma eruptions that are triggered from magnetic sites embedded in sunspot penumbral regions. Multiple trigger mechanisms for recurrent jets are under debate. Vector magnetic field data from SDO-HMI observations are used to analyze a prolific photospheric configuration, identified in extreme ultraviolet observations as a `Coronal Geyser', that triggered a set of at least 10 recurrent solar active region jets.  


We focus on interpreting the magnetic fields of small-scale flaring sites aiming to understand the processes that govern recurrent jet eruptions. We perform a custom reprocessing of the SDO-HMI products, including disambiguation and uncertainty estimation.  We scrutinized the configuration and dynamics of the photospheric magnetic structures. The magnetic configuration is described via the analysis of the photospheric magnetic vertical fields, to identify the process is responsible for driving the jet eruptions.  

We report that the two widely debated magnetic trigger processes, namely magnetic flux cancellation and magnetic flux emergence, appear to be responsible on a case by case basis for generating each eruption in our set.  We find that 4 of 10 jets were  due to flux cancellation while the rest were clearly not, and were more likely due to flux emergence.
\end{abstract}


\section{Introduction}
\label{sec-intro}

This work aims to shed light on the possible triggers of active region coronal jets (AR jets). Coronal jets are defined as collimated eruptions that are correlated to local microflaring footpoints, and are connected to the open corona, that usually escape into the inner heliosphere \citep[e.g. ][and references within]{wang1998,stcyr1997}.  Furthermore, \citet{shimojo1996,shimojo2000} studied X-Ray jets originating in active regions, quiet sun, and coronal holes, finding that $\sim$68\% of the jets appear in or near active regions. We study one set of penumbral AR jets in extreme ultraviolet (EUV) observations. The AR jets were recurrent and in part appeared as homologous in EUV observations. By recurrent we understand that all jet eruptions originate via repeated reconnection at the same EUV footpoint, and by homologous we refer to recurrent jets as having similar dimensions and behavior in EUV wavelengths. Our jet observations follow the above  defined characteristics.  A comprehensive review on coronal jets is presented by \citet{raouafi2016}.

The detailed physical processes that govern such small-scale eruptive processes are not yet fully understood.  Our initial supposition involves discerning if recurrent or homologous jets are the result of a series of self-similar reconnection events involving an arched structure (possible micro-filament) that is subjected to energy storage and release mechanisms? Alternatively, can recurrent jets result from a chaotic behavior of lower atmosphere features that may interact with a local (quasi)stable structure (possible micro-filament)? 

Theoretical works provide hints towards disentangling the nature of the erupting structure. AR jets were first simulated in 3D by \citet{moreno2008} who modeled a sheared flux rope interacting with a tilted magnetic field configuration. Additional simulations of a small active region magnetic field were performed by \citet{archontis2010} framing the eruptions inside the emerging flux scenario \citep[see review; ][]{cheung2014}. Further input was provided by \citet{pariat2009,pariat2010} who found that (at least) two distinct regimes of reconnection were manifesting. Furthermore, \citet{moreno2013} modeled a set of recurring jet ejections that resemble mini-CMEs suggesting that their physical properties may resemble blowout jets as described by the \citet[][]{moore2010} dichotomy. The authors found that the eruptions were not in fact homologous. Similar results were reported by \citet{archontis2013} who managed to reproduce distinct standard and blowout eruptions. Standard and blowout jets are described by \citet{moore2010}.  In a series of recent modeling works, \citet{pariat2015,pariat2016,wyper2018} proposed a `breakout' model of solar jets. This model proposes an erupting unstable micro-filament embedded in an open field and appears to be compatible with the \citet{sterling2015} hypothesis that discusses links between blowout jets and micro-filaments.  \citet{panesar2016} present comparable observational results of a set of very energetic recurrent jets that originated from the same arched structure, which the authors identified as a micro-filament footpoint.

Two scenarios are extensively proposed as triggers of coronal jet eruptions, possibly via micro-filament eruptions. A wealth of studies have considered flux emergence in the context of jet eruptions. Observational \citep[e.g. ][]{Chandra2015,li2015} and theoretical  \citep[e.g.][]{archontis2010,lee2015} studies have shown that an emerging flux can destabilize pre-existing magnetic structures. By combining modeling and observations, \citet{cheung2015} analyzed 4 jets originating from a pore embedded in the interior of a supergranule to obtain data-driven simulations that show that the emergence of magnetic field structures in the vicinity of a pore are compatible with recurrent jet formation. A recent case study, presented by \citet{sakaue2017,sakaue2018} showed that such flux emergence plays a significant role in producing one EUV AR jet.  

An alternative explanation for generating small scale AR jets is offered by the flux cancellation scenario, usually occurring in decaying active regions.  \citet{martin1985} and \citet[][sec. 4]{vandrielgesztelyi2015} present a scenario involving moving magnetic features \citep[MMFs;][]{harvey1973}. Small MMFs stream away from AR penumbras and coalesce with satellite spots \citep{leka1994} and were revealed to be involved in the production of X-ray jets and H$\alpha$ surges \citep{canfield1996}.  \citet{jiang2007,chen2008}, and \citet{chenh2015} showed that converging magnetic fluxes of different polarities appeared to cancel with each other at the base of jet eruptions. Flux cancellation is also backed up by the observational study of magnetic moving features  of \citet{chen2015}.  In a series of papers, \citet{sterling2016,sterling2017} analyzed sets of recurring AR jets and discussed potential trigger and emission mechanisms, concluding that magnetic flux cancellation should be considered a fundamental process for AR jet production, where the mechanism is compatible with a blowout/micro-filament eruptions. 

In this work, recurrent EUV jets are scrutinized, aiming to uncover which magnetic trigger mechanism (e.g. either flux emergence or flux cancellation) generates a series of eruptions. We analyze magnetic field observations linked to recurring EUV coronal jets that manifest for multiple days. Due to data and geometry constraints we present 10 recurring jets that were observed over 24 hours. To our knowledge, this is the first study that provides a long temporal analysis of jet recurrence, identifying a high number of jets erupting from the same site. Guided by this long lasting recurrence, we call the presumed stable EUV source a 'coronal geyser'.  

Starting from the EUV perspective, we defined \textbf{Coronal Geysers} ( see \citealt{paraschiv2018} and \citealt{paraschiv2019}) as small-scale penumbral active region structures that have long lifetimes, an open field coronal connectivity, and are prolific generators of recurrent jet eruptions. In the complementary works we address whether coronal geysers are sources of particle acceleration and radio bursts, establish that they are classified  from  an  energetic  point  of  view  as  impulsive  micro-flare  sites, can contain filamentary structures, and can be subject to helicity conservation. Here we examine the roots of one geyser structure by analyzing its complex photospheric magnetic configurations. 

The work is structured as follows: Sec. \ref{sec-obs} presents the geyser observations and instrumentation used. As the observational interpretation of lower strength magnetic fields is not a straightforward process, sec. \ref{sec-meth} describes the interpretation of uncertainties and the implementation of a custom disambiguation of the 180$^{\circ}$ uncertainty inherent in vector magnetic field measurements. Sec. \ref{sec-mmf} discusses the lower photospheric manifestations that we find to be responsible for triggering coronal jets and describes the applicability of the flux emergence or the flux cancellation scenarios for each individual eruption. The implications are debated in sec. \ref{sec-disc}.  

\section{Observations and Instrumentation}\label{sec-obs}
\subsection{The penumbral AR11302 geyser site}\label{sec-obs:sub-obs}

A recurrent jet site was observed in EUV at the south-eastern periphery of AR11302. We detected numerous EUV jets associated with an unique footpoint, identified as the geyser, during multiple days of AR11302's near-side crossing.  We analyzed 10 jet eruptions occurring on 25 Sep 2011. To highlight the jet activity, we consider the multiwavelength observations in the EUV channels of Solar Dynamics Observatory, \citep[SDO;][]{pesnell2012} Atmospheric Imaging Assembly \citep[AIA;][]{lemen2012} and the Solar Terrestrial Relations Observatory \citep[STEREO;][]{kaiser2008}, Extreme Ultraviolet Imager \citep[EUVI;][]{wuelser2004}. The AIA and EUVI instruments showed the geyser's activity from different vantage points of the SDO and STEREO spacecrafts. The AIA EUV imager observes the ultraviolet sun with a spatial sampling of $\sim 0.6'' \cdot$ pix$^{-1}$ and a temporal cadence of 12 s. The STEREO EUVI imager has a spatial sampling of $\sim 1.6'' \cdot$ pix$^{-1}$ and a temporal cadence of 5 min. The raw data were extracted via the JSOC and SECCHI pipelines and calibrated locally to level 1.5. The data was further processed using the Solarsoft \citep[SSWIDL;][]{freeland1998} package and additional corrections were applied, e.g. pointing, co-alignment, re-spiking, aia\_prep corrections, secchi\_prep, etc. 
  
\begin{figure}[!ht]
\centering
\includegraphics[width=0.79\linewidth]{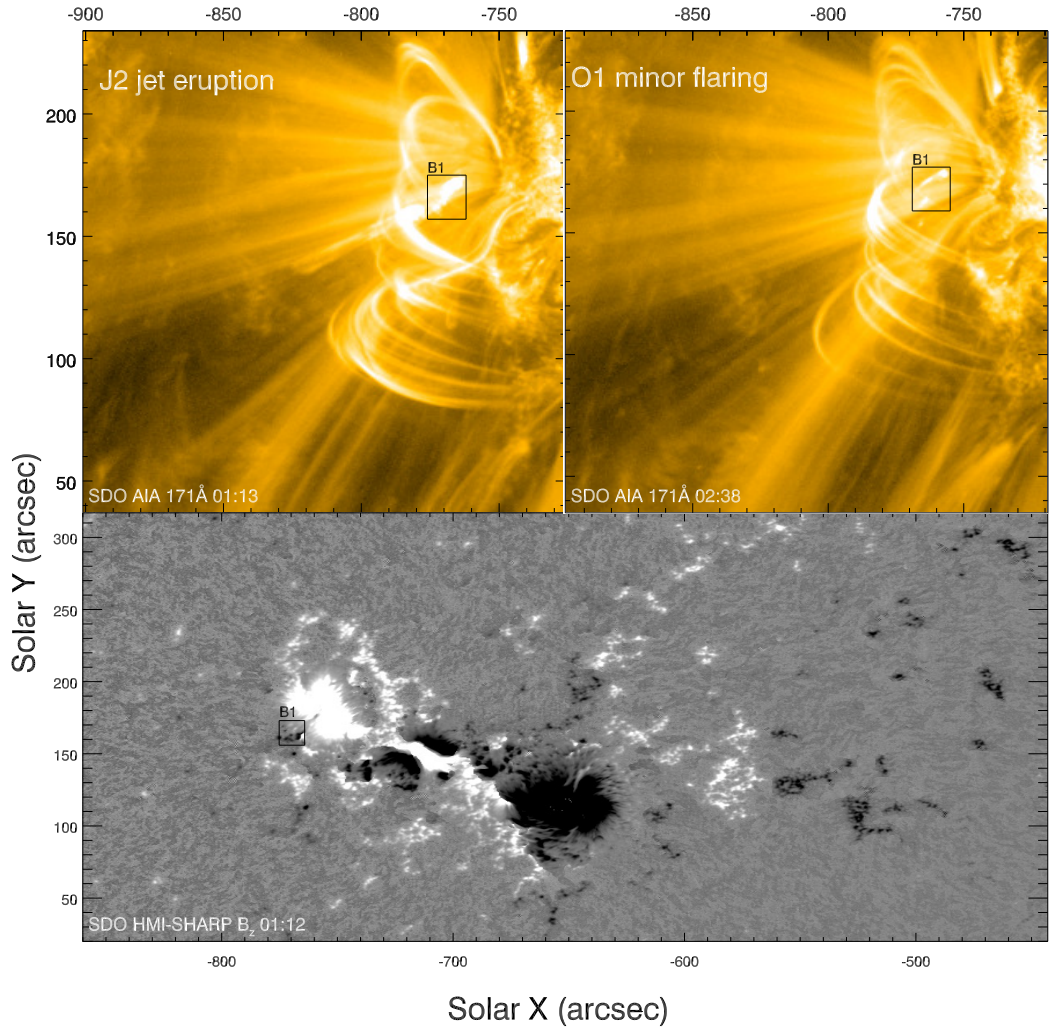}
\caption{Context view of EUV AR jets, minor flaring and a co-aligned region of interest in AIA 171\AA{} and HMI-SHARP vertical $B_z$ magnetic flux. \textbf{Top panel:} A $13\arcsec\,\times\,17\arcsec$ region (B1) is centered around the Geyser site (jet reconnection site in EUV) using SDO's AIA-171\AA{} filter. Two time instances are presented, the J2 jet and the O1 minor flaring event. In the case of O1-O7 events, no jet outflow could be observed in any AIA EUV filter. \textbf{Bottom panel:} Context view of the AR11302 vertical component $B_z$ of photospheric flux at 01:12TAI. The modest size penumbral B1 region is highlighted at approximately [-750$\arcsec$,175$\arcsec$].}
\label{fig-b1-bz-aia}
\end{figure}  
  
 
 All recurrent jets (identified as J1, J2, ... J10) originated from the geyser structure. A 13$''$ x 17$''$  region, labeled B1, was identified and selected corresponding to the onset location of all observed recurrent jets. Observing a jet is summarized as the detection of sudden brightness increase in EUV at the geyser location followed by the observation of collimated erupting material. We also observed that minor flaring events (O1-O7) are spatially correlated with the geyser location, and manifested individually or in groups, although no erupting material could be detected in any of the AIA EUV bandpasses, hence distinguishing them from jets. See the visual comparison between jets and minor flaring in Fig. \ref{fig-b1-bz-aia} (top). The Fig. \ref{fig-b1-bz-aia} (bottom) context  magnetogram presents the location of the B1 region, highlighting its modest size with respect to AR11302. From the EUV perspective, all  jets followed the same propagation direction, erupting along a presumably open magnetic structure (see change in topology in right panel of Fig. \ref{fig-potheight}).

\begin{figure}[!hb]
\centering
\includegraphics[width=0.98\linewidth]{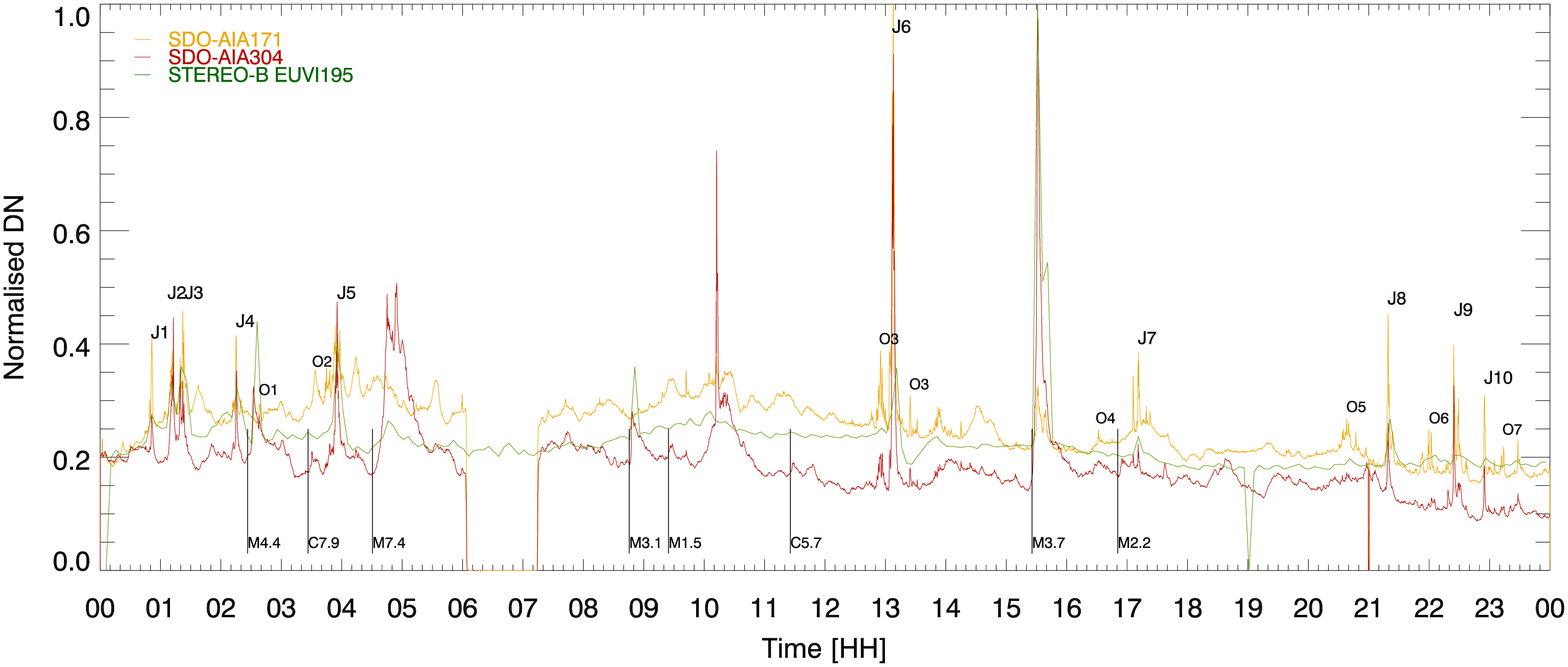}
\hspace{-0.4cm}\includegraphics[width=0.99\linewidth]{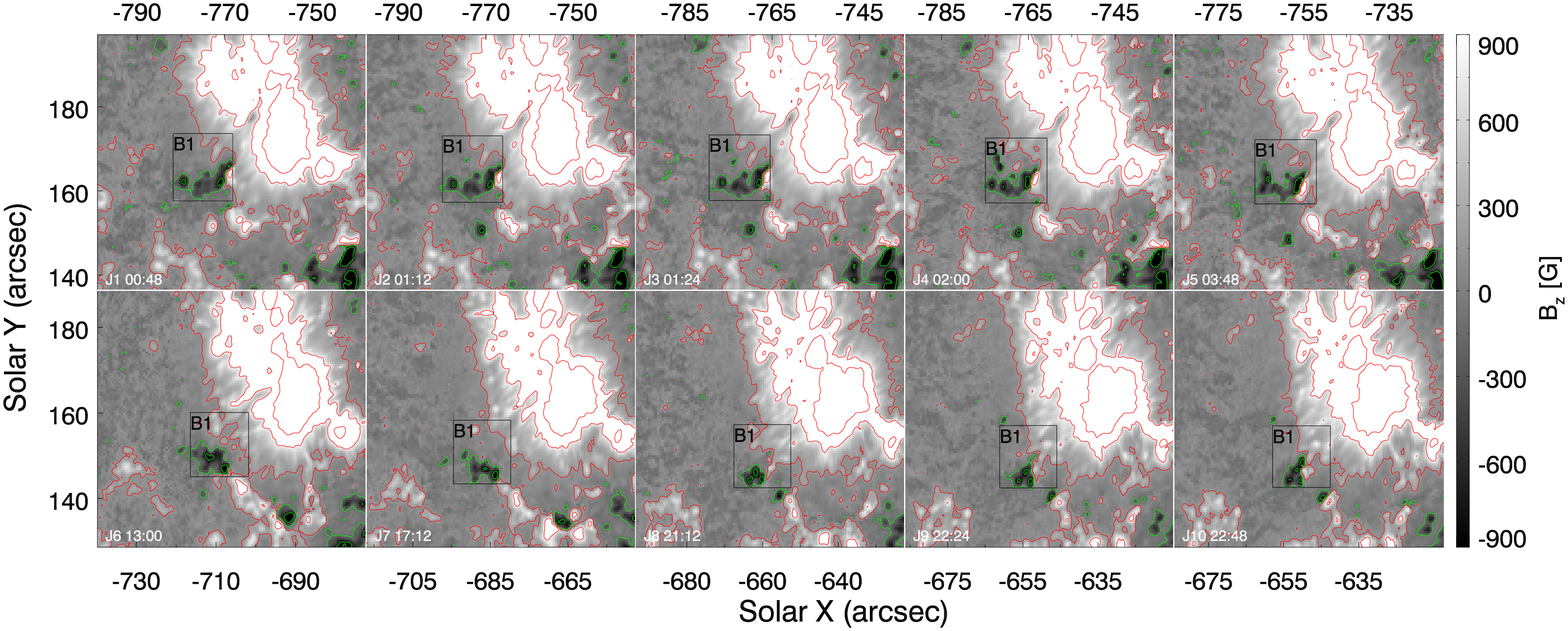}
\caption{\textbf{Top panel:} Geyser structure 24 h region tracking of a  $13\arcsec\,\times\,17\arcsec$ region (B1) centered around the Geyser site (jet reconnection site) using SDO's AIA-171\AA{}, AIA-304\AA{}, and STEREO-B EUVI-195\AA{} filters. Two gaps are present in the AIA data timeseries at 06:03-07:12 and 21:00-21:02, and one in the STEREO-B series at 19:00-19:10. Individual jet events are labeled J1-J10 and other peculiar geyser flaring, that do not exhibit any distinguishable jet emission are labeled O1-O7 (see sec. \ref{sec-obs:sub-obs}). Major AR11302 flaring events that can potentially interact with the integrating region are also illustrated.  \textbf{Bottom panel:} Maps of the vertical component $B_z$ of photospheric magnetic fields depicted before (up to 12 minutes due to cadence)  each individual jet eruption. The B1 region is highlighted. Contours at $\pm160G,~\pm800G,~\pm1500G $ levels, for positive (red) and negative (green) fluxes are shown in order to enhance visibility of small MMFs, emerging flux, and potential inversion lines related to jet activity.}
\label{fig-bzcont-aia}
\end{figure}

 To demonstrate the spatial uniqueness of the structure, we employ a timeseries analysis of the EUV data using the B1 region. We apply an in-house developed full cadence spatial tracking procedure to the AIA-171\AA{}, AIA-304\AA{}, and STEREO-B EUVI 195\AA{} data. The procedure computes a timeseries of the normalized pixel-averaged intensity inside the B1 region, while accounting for solar projection and rotation effects that manifest when dealing with a long temporal tracking. The resulting intensity light-curves are shown in the top panel of  Fig. \ref{fig-bzcont-aia}. Jet eruptions J1-J10 and minor flaring events O1-O7 originate from the unique site. Major AR11302 flaring events that can potentially interact with the B1 integrating region are also illustrated.  As can be seen, these are not directly related to any of the jets or minor flaring events. An adaptation of this EUV region tracking procedure that instead tracks two pseudo-slits, S1 and S2, is used to highlight and analyze the magnetic features inside the B1 region (see sec. \ref{sec-mmf}). Both the slit and region tracking procedures along with usage examples are available in the associated Harvard Dataverse repository \citep[][V1]{paraschiv2019-data}.   

\subsection{Vector magnetic field observations of the AR11302 Geyser Site}\label{sec-obs:sub-hmi}

When probing the photospheric magnetic fields of AR11302, it is observed that the underlying configuration plays a crucial role in continuously triggering the coronal jets. Our data analysis is based on the interpretation of the penumbral magnetic fields.  Both SDO Helioseismic and Magnetic Imager \citep[HMI; ][]{scherrer2012} and  the Hinode \citep{kosugi2007} Spectro-Polarimeter \citep[SP][]{tsuneta2008} observations of AR11302 were available on 25 Sep 2011, although the only available Hinode-SP AR11302 scan has a datagap over our region of interest, preventing any analysis possibility. Thus, only the HMI magnetogram data is utilized in order to recover the lower atmosphere features linked to the geyser eruption site.  


HMI imaging system has a 1$\arcsec$ optical resolution and samples 6 wavelength positions centered on the \ion{Fe}{1} 6173.34\AA{} line.  The data are processed according to \citet{hoeksema2014} and \citet{bobra2014}. We employ the \emph{hmi.sharp\_720s} dataseries magnetograms available through JSOC. 

During our 24 h observing period, the geyser structure manifests inside the B1 region as depicted in Fig. \ref{fig-bzcont-aia}. The structure is located close to center latitudes ($\sim140\arcsec \rightarrow180\arcsec$) but at significant high longitudes ($\sim-790\arcsec\rightarrow -650\arcsec$). Vector magnetograms of structures far from disk center are considered unreliable \citep[e.g.][]{hoeksema2014,falconer2016,bobra2014} and additional steps are required for an accurate analysis.  Due to the high longitude observation, we perform a custom interpretation of the HMI data by taking into account: the projection effects, the local measurement uncertainties, and the significant polarity ambiguity issues. We note that the complementary dataseries, the HMI line of sight magnetograms,  are compromised by limb effects.     
\vspace{1cm}

\section{Methods: Interpreting HMI small-scale observations}\label{sec-meth}
\subsection{HMI coordinate transform }\label{sec-meth:sub-coord}

\indent\indent Instrumental and physical issues need to be addressed when performing photospheric vector magnetic field data analysis.  \citet{hoeksema2014} presents an extensive overview of assumptions, issues, and uncertainties that arise when creating the SDO-HMI vector maps; we briefly discuss only those related to our observations.

 The data series \emph{hmi.sharp\_720s} provides three main magnetic field products: the total field strength ($\vec{B}$), the inclination ($\gamma$) and the azimuth angle ($\phi$). The  inclination is calculated with respect to SDO's line of sight since we are dealing with image based coordinates.  This aspect becomes important as our target active region is close to the disk center, hence significant projection effects exist.
 
 The HMI data are subject to significant uncertainties when recovering low and intermediate field magnetic fluxes. The JSOC  \emph{hmi.sharp\_720s} dataseries pipeline  does not perform automatic remapping to solar tangential magnetic field vectors. We note that automatic remapping is available for the \emph{hmi.sharp.CEA\_720s} series that we do not use here due to the more convoluted co-alignment issues with the EUV data.  We transform the magnetic vectors into a heliographic-tangential system and projected them on a new solar plane using the coordinate transforms given by \citet{gary1990}. We note that the plots axis are kept in disk-arcsecond coordinates to easily associate to co-aligned observations (e.g. SDO-AIA, etc.).  

\subsection{Local HMI uncertainty estimation}\label{sec-meth:sub-uncert}

\indent\indent Once the heliographic field components were calculated, we investigate the uncertainties correspondent to our observation. The solar disk position of the SHARP region is far from the disk center. In particular, the inclination angle ($\gamma$) component is noisy leading to higher than standard vector HMI uncertainties. 

HMI penumbral magnetic flux is generally of medium strength (250 - 700 G), being significant but not very strong when compared with the quiet-sun disk center background levels ($<$100 G). Following a standard error propagation we assume that the $B$, $\gamma$, and $\phi$ quantities are independent variables and derive the uncertainty maps corresponding to the image projected vectors. We further subject the maps to the heliospheric projection transform described in sec. \ref{sec-meth:sub-coord}.  

Furthermore, we performe a custom estimation of observational detection limits. Two statistical constraints that could influence the interpretation of the observations are investigated.

\textbf{First,} in order to understand the influence of solar position in evaluating systematic uncertainties, we derive the detection limits of the background values of transverse fields ($B_t$). This evaluation is important as fields in or near sunspot penumbra tend to have strong horizontal components. As AR11302 moves towards a disk central position, we consider the first two hours of observation, 00:00TAI to 02:00TAI, to get an upper limit on the actual systematic noise. 

\begin{figure}[t]
\centering
\includegraphics[width=0.49\linewidth]{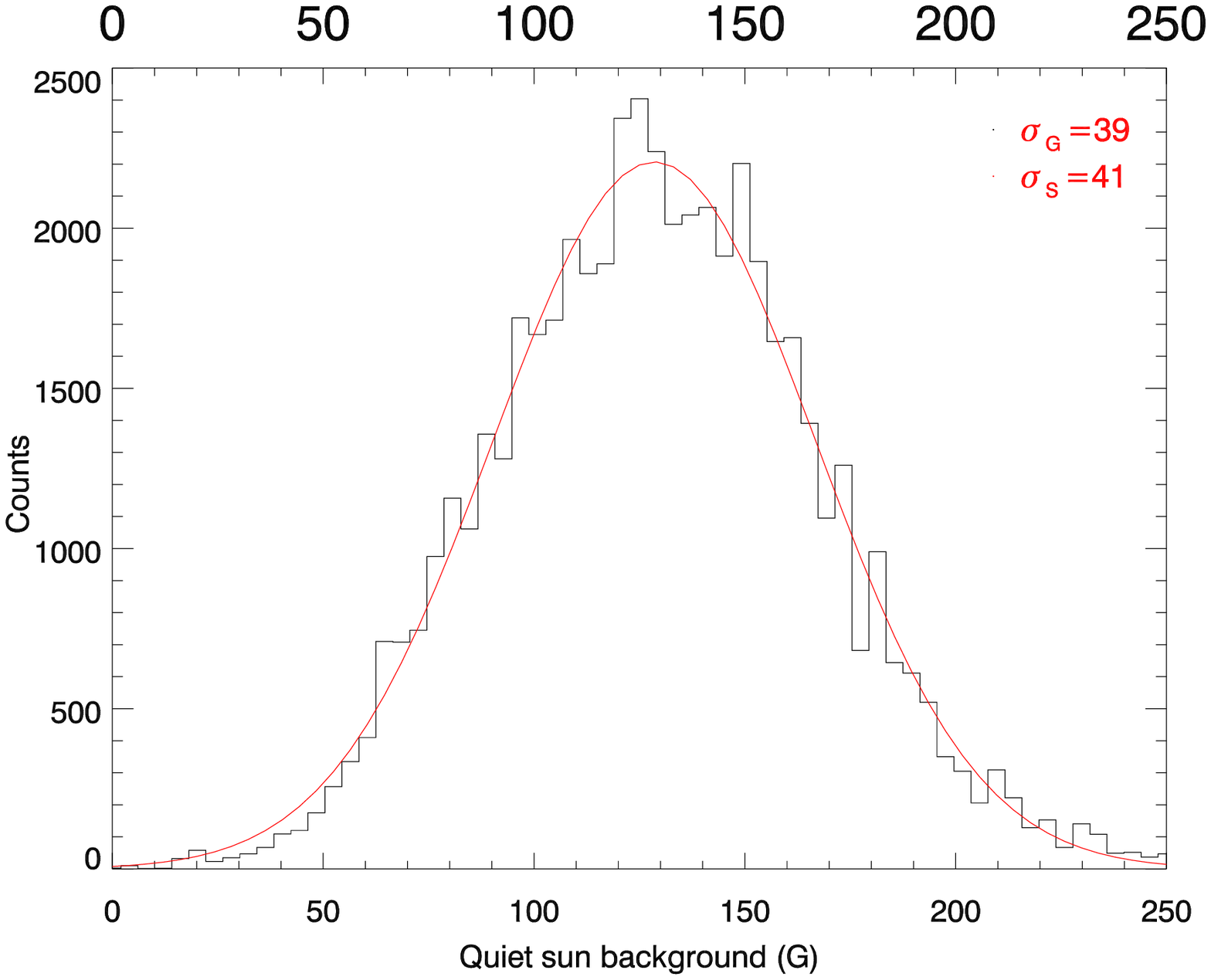}
\includegraphics[width=0.49\linewidth]{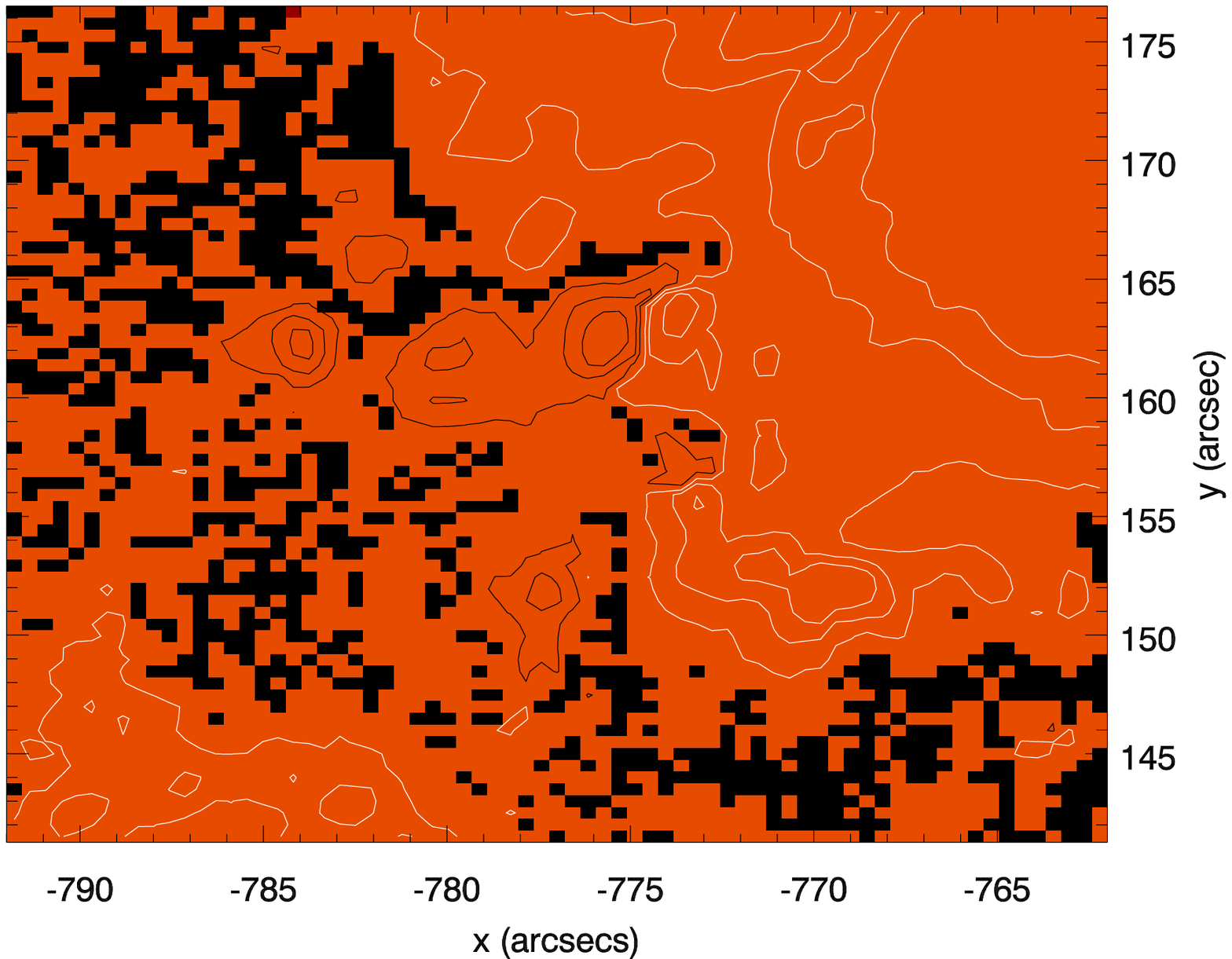}
\caption{Estimation of the detection limits and pixel uncertainties of SHARP region 892 (AR11302) for a two hour interval between 00:00TAI and 02:00TAI on 25.09.2011. We evaluate these uncertainties assuming that these are the higher limit of errors resulted from projection effects, as the AR is moving towards a disk central position. \textbf{Left Panel:} Histogram of the transverse fields ($B_t$) noise values near the position of the geyser. The standard deviation $\sigma_S$ along with a fitted Gaussian standard deviation $\sigma_G$ are calculated. \textbf{Right Panel:} Example (01:24TAI) of a custom confidence map of the photospheric geyser region. The relative uncertainty of the  heliospheric vertical magnetic fields ($B_z$) is computed using the eq. \ref{eq-berr} validation criterion. The contours represent fields of strength [$\pm$200 G, $\pm$500 G, $\pm$800 G] of positive (white) and negative (black) polarity and the black patches represent uncertain areas.}
\label{fig-noise}
\end{figure}

We present the resulting $B_t$ background noise distribution in fig. \ref{fig-noise} (left). We selected and tracked regions that were qualitatively void of any significant magnetic structure, located in close proximity ($<$ 25$\arcsec$ towards south-east) to the geyser (B1 region). 99.2\% of the selected pixels in the noise sampling region had $B_t$ $<$ 250 G during the 2 hour sampling interval. Scarce pixels (0.8\%) with $B_t$ between 250 G and 460 G were identified, but were not shown in the plot. However, they were included when computing both types of deviations.

A mean noise threshold value of $\bar{B_{t}}\sim$130 G is found. Similar values for the standard error propagation deviations, $\sigma_S=41$ G, and fitted Gaussian deviations, $\sigma_G=39$ G, are obtained using fig. \ref{fig-noise} (left). 

As hypothesized, the noise level is particularly high. \citet{hanson2015} found noise values of 60 G in the quiet sun patches neighboring an AR close to disk center.  In our case, due to both the medium strength fields involved in jet eruptions and high $\bar{B_{t}}$ background level, we have chosen our $B_t$ detection limits to be accurate to at least $3\sigma$ (e.g. $B_t >250$ G). 

\textbf{Second,} we computed the confidence map of the photospheric fields corresponding to our geyser structure (see fig. \ref{fig-noise}, right).  The map represents the relative uncertainty of the heliospheric vertical magnetic fields ($B_z$), computed using  the  
  \begin{equation}
\frac{B_z^2}{B^2_{z-err}}>1, 
\label{eq-berr}
\end{equation}   
 validation criterion employed by \citet{hanson2015}. One frame, 01:24TAI, close to the start of the observations is chosen with the intent to sample a measure of the highest magnetic uncertainties. We  checked the confidence maps for multiple time instances finding similar results. 

  The fig. \ref{fig-noise} (right)  map is dual tone, where the red areas represent usable pixels and the black patches represent uncertain areas that are dominated by noise. We overplotted contours of vertical fields of strength [$\pm$200 G, $\pm$500 G, $\pm$800 G] of positive (white) and negative (black) polarity respectively.   This shows that the MMF's, small dipole emergences, and the stronger pore structures (described in sec \ref{sec-mmf:sub-ingr}) are not noise dominated.  We tracked the the vertical field component ($B_z$) uncertainties of the (above-noise) magnetic fields of our photospheric region of interest. The uncertainties varied on average from typical values of $\pm$90 G to $\pm$50 G during the 24 h of observations as the region rotated towards disk center. 

\subsection{Custom HMI disambiguation }\label{sec-meth:sub-diss}
 
\indent\indent A 180$^{\circ}$ ambiguity in the azimuth of the recovered photospheric magnetic fields exists \citep{harvey1969} and multiple solutions were developed to solve this issue \citep[see review; ][]{metcalf2006}. The HMI pipeline implements the minimum energy ambiguity resolution method  \citep[ME0:][]{metcalf1994,leka2009}. As our interest lies in obtaining the most accurate representation of low and medium strength magnetic fields, we focus on how the JSOC pipeline implements ME0 \citep[see ][]{hoeksema2014, leka2009} and devise a custom disambiguation setup tailored to our case using the AMBIG-ME0 code\footnote{AMBIG-ME0 code available at \url{www.nwra.com/AMBIG}}.


The detection limit of our HMI transverse fields were found in sec. \ref{sec-meth:sub-uncert} to be  $B_t>$250 G. We devise a stable custom parameter scheme which is suited to our goals. Our AMBIG runs are set to perform a full SHARP region minimization for the fields which are above the quiet sun noise level (athresh=120) and subjected patches that are under the 3$\sigma$ strength (bthresh=250) to nearest neighbor smoothing. The number of times each pixel can be visited at each annealing iteration was substantially increased (neq=200) to allow for a greater number of small patch flips. We set a higher starting temperature (tfac0=2.4) and a significantly slower annealing (tfactr=0.998) conditions, as it creates a better chance of converging to the global minimum.
 
We tested the stability of the disambiguation results by performing 5 individual runs using different initial randomization seeds (iseed). We verified that the geyser region had no discernable patches showing checkerboard patterns, lack of smoothness,  and no unstable polarity flips occurred  during our observation. Using this setup we have found that no issues arose for magnetic fields above our 3$\sigma$ detection limits and concluded that for this particular region, the field orientation proved stable after the re-calculation of the disambiguation. 

As shown, the retrieval of magnetic vector fields outside a disk central position and the small-scale nature of the magnetic fields require a cautious interpretation. All uncertain pixels that were found to be unreliable due to projection, uncertainties or disambiguation issues were excluded from further analysis. To provide an straightforward  way of replicating the results, we refer the reader to the fully processed 24 h data cubes including disambiguated maps and uncertainty estimations along with the slit procedure and usage examples. Data and code can be found in the associated Harvard Dataverse repository \citep[][V1]{paraschiv2019-data}.  \newpage

\section{Results: Magnetic features at the periphery of AR11302}\label{sec-mmf}
\subsection{Geyser site magnetic ingredients}\label{sec-mmf:sub-ingr}

In order to understand why this region was so efficient in generating recurring jets, we investigated the photospheric rapid motions of observed magnetic polarities involved in generating the detected EUV jets. We initially assumed that all the 10 jet eruptions could be correlated to either a flux cancellation or a flux emergence scenario. We now  discuss the magnetic configuration of the EUV observed geyser structure in order to identify the best suited scenario. 

\begin{figure}[!b]
\centering
\includegraphics[width=0.37\linewidth]{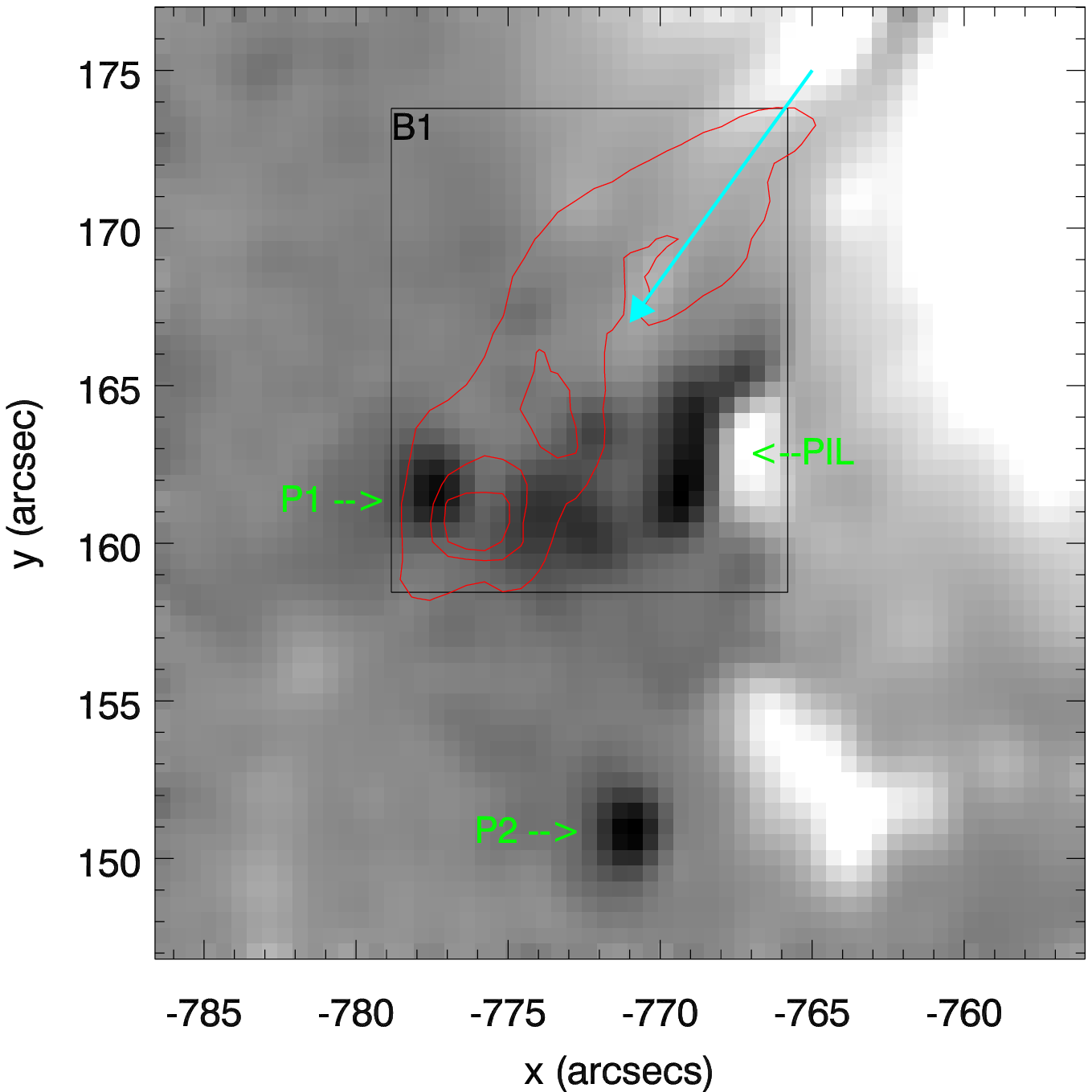}
\includegraphics[width=0.62\linewidth,trim={0cm 2.cm 0cm 3.cm}, clip]{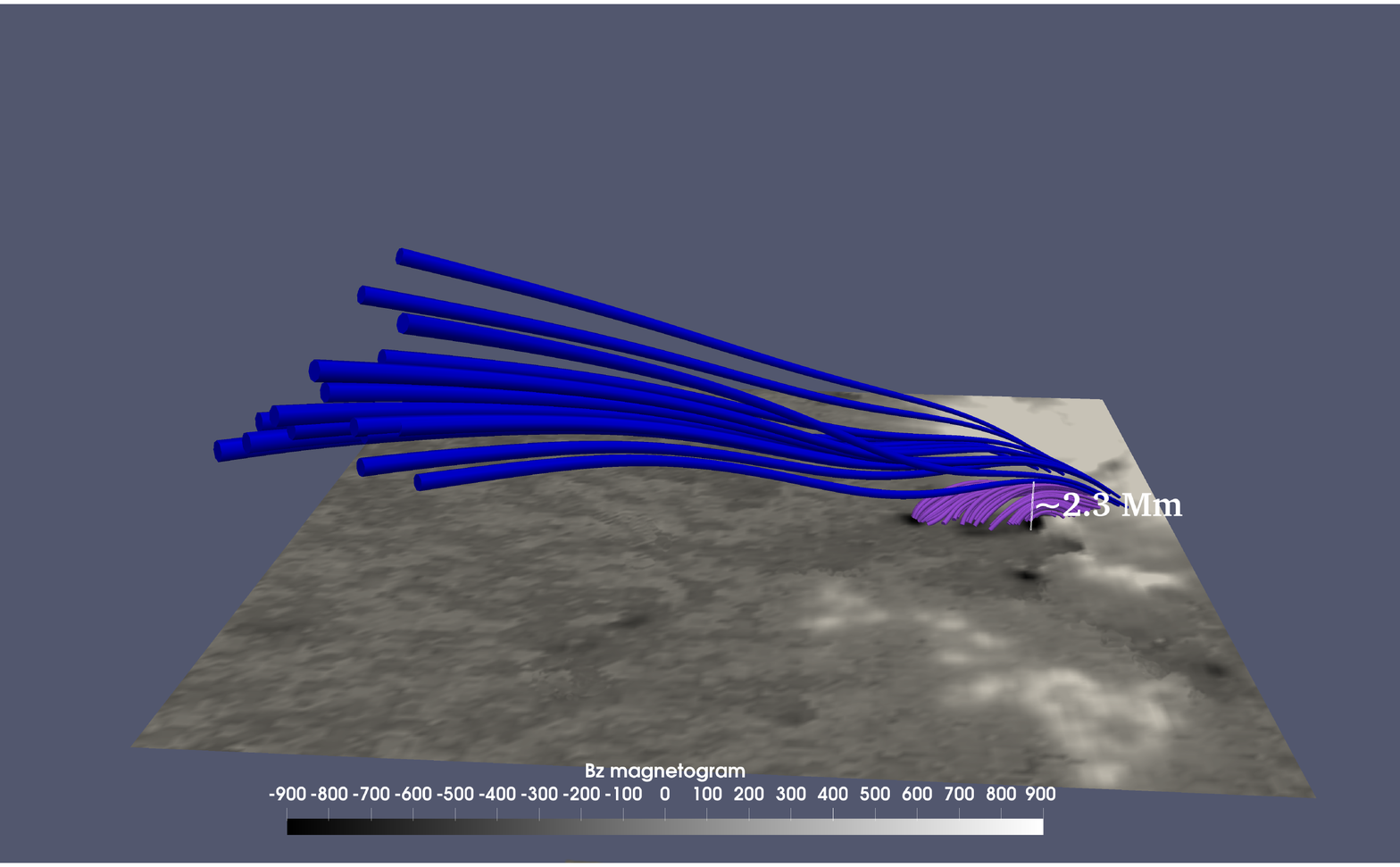}
\caption{Magnetic moat at the periphery of AR11302 at 01:12TAI. \textbf{Left Panel:} Vertical $B_z$ magnetogram snapshot of the penumbral moat at the south-east of AR11302. The P1 and P2 pores, and the dipole polarity along with its potential inversion line (PIL) are highlighted. The turquoise arrow shows the general outflow direction of MMFs. The B1 region is drawn  along with a contour (red) of the flaring associated to the J2 jet, as seen in the `hotter' AIA-94\AA{} filter at 01:13TAI. \textbf{Right Panel:} A potential field extrapolation is constructed using the \citet{alissandrakis1981} method from the $B_z$ magnetogram. We show here the penumbral canopy (blue) that is situated above our region of interest (purple). }
\label{fig-potheight}
\end{figure}

A set of long lived vertical field structures, emerging fields, and moving polarities manifested in our region of interest embedded in AR11032's penumbral moat. The magnetic data was recovered using the methods and assumptions described in sec. \ref{sec-meth}. If not mentioned otherwise, we describe the position and field strength of structures in reference to the starting time of observations (00:00TAI).

\textbf{Moat:} The eastern sunspot of AR11302 has a positive magnetic polarity and is surrounded by a highly dynamic moat, characterized by continuous magnetic flux emergences and cancellations. The moat region is continuously evolving during the surveyed 24 hours (fig. \ref{fig-bzcont-aia}, bottom). The radially outward propagation, and photospheric motions inside the moat are associated with the decay phase of the sunspot development.

\textbf{Pores:} These are stronger flux concentrations and do not move over short times-scales when compared to MMFs.  Fig. \ref{fig-potheight} (left) shows the main pore P1 of peak initial strength $\sim-900\pm80 G$ that is linked to the jets' generation. P1 initially appears at $(X,Y)\sim(-780\arcsec,160\arcsec)$. A second pore (P2) is observed at about 10 Mm distance at $(X,Y)\sim(-775\arcsec,150\arcsec)$ but was found to not be involved in any of the jet eruptions or minor flares. 

\textbf{Dipole polarity and potential inversion line:} A notable structure in Fig. \ref{fig-potheight} (left)  is the small-scale polarity inversion line (PIL) present along a dipole structure. The dipole appears at position $(X,Y)\sim(-770\arcsec,165\arcsec)$, is of initial $B_z$ peak strength $\sim\pm$800$\pm$80 and tends to slowly drift southward over the observed 24 hours (fig. \ref{fig-bzcont-aia}, bottom). This movement is much slower than the motion of the positive MMFs that are dragged around by the moat flow. During this time the PIL-dipole morphology is drastically changing. Regardless, the PIL is clearly distinguished during the 24 h observations.


\textbf{Penumbral canopy}: Above the highly dynamic moat, the hosting penumbra sustains a steady structure, part of which is recognized as open magnetic flux  that significantly contributes to the coronal connection of the underlying moat features.  The magnetic canopy is also rooted in the penumbra of the sunspot, where these magnetic structures rise and arch above the magnetic moat, then radially extend towards the outer heliosphere. 
In order to reveal the separation between the low lying moat fields and the open canopy structure we constructed potential field extrapolation from the $B_z$ magnetograms using the \citet{alissandrakis1981} method. Qualitatively, the potential extrapolation shown in Fig. \ref{fig-potheight} (right) indicates that fields that arch to heights $>2Mm$ above the B1 region become part of this canopy  structure (blue). The observed jet eruptions are guided along the direction of this canopy (see EUV jet in top panel of  Fig. \ref{fig-b1-bz-aia}). The extrapolations are constructed only for this purpose and are not utilized to assess the trigger scenarios of the jets.

\textbf{Magnetic moving features:} The vector magnetograms often reveal moving magnetic features (MMFs) of modest sizes ($\sim0.5$ Mm). The general propagation (advection) direction for these MMFs is shown by the turquoise arrow in Fig. \ref{fig-potheight} (left). They carry flux of the same polarity (positive) as the active region and originate in its penumbra, streaming outward in the direction of the negative pore with typical speeds in the order of 0.4-0.8 km s$^{-1}$, eventually cancelling with the main pore. Inflowing and opposite polarity MMFs are not detected for our region of interest.

\textbf{Newly emergent magnetic fields:} The locations of the flux emergence areas are co-spatial and co-temporal with the EUV hot filament observed during flaring. This suggests that the small scale dipoles allow for efficient rising of low lying loops, that later interact with the above penumbral canopy fields, triggering the release of jet eruptions observed in the high corona and heliosphere.

The periphery of AR11302 was highly dynamic and active, and with the use of our tailored methods, allowed us to study the photospheric magnetic field evolution leading to each jet eruption.

\subsection{Jets generated by magnetic flux cancellation}\label{sec-mmf:sub-canc}

\indent\indent  Noticeable are a few MMFs of positive polarity that were identified inside the penumbral moat, moving southward, with speeds on the order of $0.4-0.7\pm0.2 \text{ km s}^{-1}$. The positive MMF's migrate towards the S-E edge of the penumbra, in the direction of the negative magnetic pore, eventually canceling.  We identified this as contributing towards generating our coronal jets. The motion of  polarities facilitates the rising of low lying magnetic loops, that eventually will reconnect with the above canopy. The magnetic cancellations of the tracked MMFs occur during the 720 s integration time of the SHARP data, when the jet footpoint flared and eruptions were observed in the higher cadence EUV data. 

\begin{figure}[!t]
\hspace{-0.7cm}\includegraphics[width=1.04\linewidth]{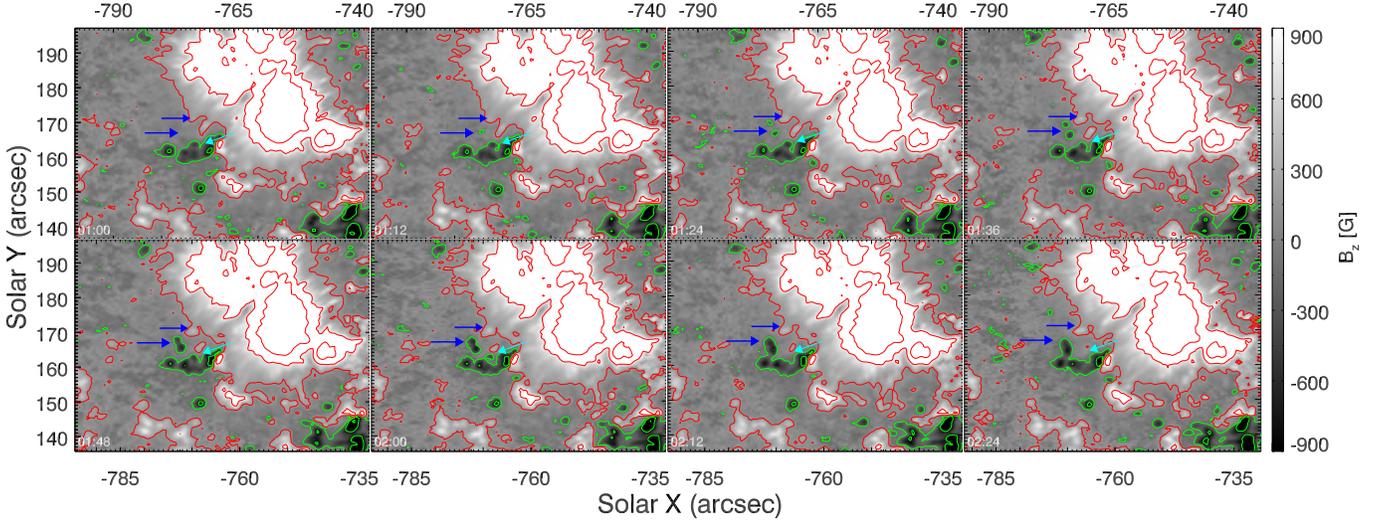}
\caption{Timeseries of the vertical component $B_z$ of photospheric fields in between 01:00TAI and 02:24TAI, covering the timespan of J2, J3, and J4 events. Contours at $\pm160G,~\pm800G,~>\pm900G $ levels, for positive (red) and negative (green) fluxes are depicted. A region where a dipole flux emergence develops during this period is shown by the two blue arrows that correspond to the emerging footpoints. The first vertical fields appear in close temporal and perfect spatial proximity with the J2-J3 events. The turquoise arrow points to a region where a positive MMF that travels outwards from the penumbra interacts with a small negative flux at the border of the pore, cancelling out during a time frame in which the J4 jet erupts in the higher EUV atmosphere. }
\label{fig-bzts}
\end{figure}

Let's consider the J4 event (Fig. \ref{fig-bzcont-aia}). The evolution of a positive MMF up to the moment of cancellation with the pore is presented in fig. \ref{fig-bzts} (turquoise arrow).  To highlight the temporal correlation between the cancellation event and  the J4 EUV eruption, we applied a tracking slit procedure that accounts for solar rotation. This is similar to the tracking of the B1 region in EUV observations discussed in sec. \ref{sec-obs:sub-obs}. We position one such slit along the general direction of MMF movement to track the $B_z$ field component of the positive polarity MMF originating in the field penumbra at $(X,Y)=(-775\arcsec, 172\arcsec )$ during 00:00TAI - 03:00TAI. The turquoise slit S2 in Fig. \ref{fig-slit} tracks the MMF of $\sim$ 530$\pm$80 G as it moved outwards in the south-east direction with an apparent horizontal speed of $v_s=0.6$ km s$^{-1}$, as derived via a time-distance fit. The positive MMF cancels shortly after 02:12TAI HMI frame by interacting with the edge of the negative pore at $(X,Y)=(-778\arcsec,165\arcsec)$. The canceled negative flux had local $B_z\sim$ -450$\pm$70 G. The AIA-171\AA{} flaring starts at 02:14TAI and the main jet is seen at 02:17TAI. Both features dissipate afterwards. No other magnetic events occurred in any of the 02:00TAI-02:24TAI snapshots.

All jets which correlated to magnetic cancellation (see Table \ref{table-dataset}) were analyzed using the same methodology and were found to follow a similar scenario to the discussed J4 jet. We find that four events, J1, J4 (fig. \ref{fig-slit}), J6, and possibly the uncertain J7 are consistent with the flux cancellation scenario. We draw attention to the fact that the magnetic fields of the J1, J4, and J6 jets are above the discussed detection limits (see sec. \ref{sec-meth:sub-uncert}). J7 resulted from the cancellation of a very faint MMF measured to have $B_z$ strength under our imposed detection limits ($B_z< 3\sigma$) and was marked as uncertain. Although faint, this positive magnetic polarity showed stability (no polarity flips, see disambiguation solution discussed in sec. \ref{sec-meth:sub-diss} ) allowing us to track its motion. We estimate an average horizontal speed of  $v_s\approx 0.6$ km s$^{-1}$ before cancellation using time-distance slit fits. No other magnetic features could be correlated to this jet.\newpage

\subsection{Jets Generated by small-scale flux emergence}\label{sec-mmf:sub-emerg}

\indent\indent We identified multiple new small scale emerging magnetic dipoles manifesting independently in the $B_z$ magnetogram series. A few particularities discern emerging flux occurrences from magnetic cancellations, namely the emergent fields last longer and are stationary. 

\begin{figure}[!b]
\centering
\includegraphics[width=0.99\linewidth]{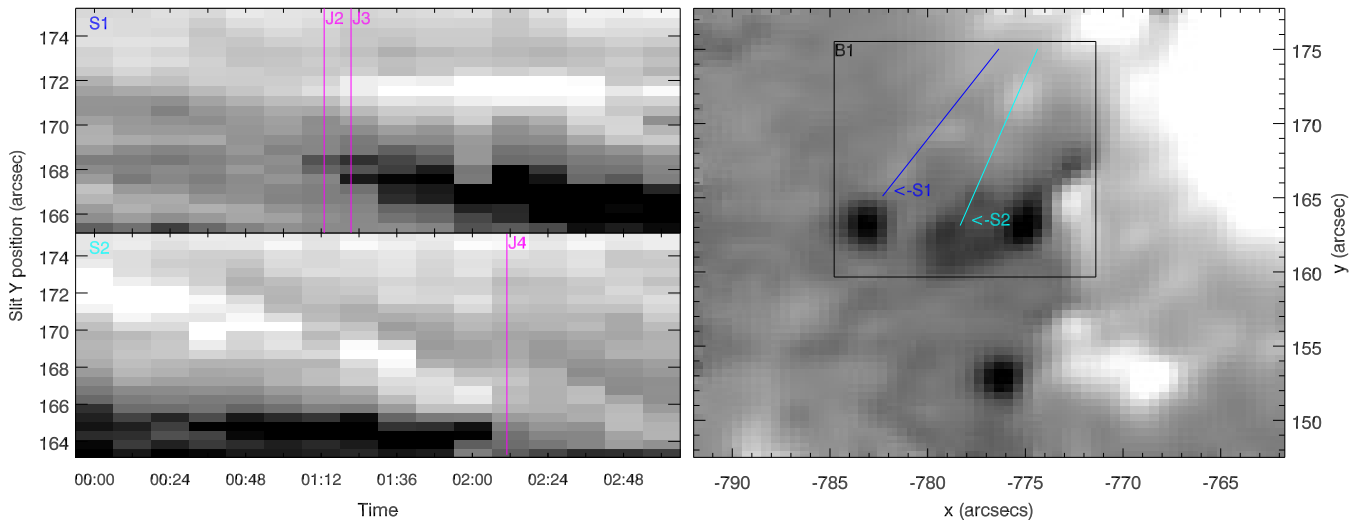}
\caption{The slit based temporal tracking of photospheric magnetic moving features and vertical flux emergence are presented.  \textbf{Right Panel:} The geyser location, outlined by the B1 region, is represented via the vertical field component $B_z$, plotted at 00:00TAI. The B1 box that was used to integrate the geyser EUV signal, is also presented.  The slits are positioned along the direction of the relevant magnetic features and their positions are temporally tracked to account for solar rotation. \textbf{Left Panel:} Slit time-distance diagrams tracking of $B_z$ at the position of the two slits. The S1 slit presents a new dipole emergence, at the 01:12TAI HMI SHARP frame, is co-temporal with the J2 and J3 events.  The AIA flaring starts at 01:13TAI and 01:19TAI. In the case of the S2 slit, we track a positive MMF that moves towards south-east with a surface speed of $v_s=0.6$ km s$^{-1}$. The feature cancels with the negative pore group in between 02:06TAI-02:18TAI (HMI SHARP frame 02:12TAI).  The AIA flaring for J4 starts at 02:09TAI.}
\label{fig-slit}
\end{figure}

A photospheric flux emergence was detected co-temporally to the J2 and J3 eruptions, which themselves could not be correlated to any cancellation of advected photospheric MMFs. Fig. \ref{fig-bzts} (blue arrows) presents the dipole emergence in the geyser region. We used slit S1 to track the region where the newly emerged dipole manifested in between 00:00TAI-03:00TAI.  Fig. \ref{fig-slit} shows the S1 slit that is tracked inside the geyser B1 region. A dipole first appeared in the 01:12TAI frame at location $(X,Y)=(-780\arcsec,167\arcsec)$. The newly emergent fields rose close to the pore which is also involved in producing J1 ($<4\arcsec$). 

We found that the J2 and J3 jet eruptions were co-temporally generated with the first observations of the newly risen dipole. In the case of subsequent short interval events, we could not distinguish changes in the emerging dipoles due to the 720s temporal data cadence. Higher data cadence (135s series) is unsuitable due to significantly higher noise levels. The AIA-171\AA{} and AIA-94\AA{} flaring starts at 01:11TAI with a maximum at 01:13TAI, when a jet is observed. A second event is recurring at 01:19TAI with a jet erupting at 01:22TAI.

The new magnetic dipole is co-spatial with the AIA-94\AA{} hot emission. For all jets, the EUV emission is described as an elongated ribbon tracing the PIL and warping towards the nearby pore. Additionally, some of the jets, namely J2, J3, J5, (flux emergence) and J6 (flux cancellation)  involved a second flaring loop (double loop flaring) that brightened during their flaring phases. This loop overlapped almost parallel to the main flaring ribbon. The detailed EUV morphology for J2, J3, and J6 can be found in \citet[][chap. 1]{paraschiv2018}.  We hypothesize here that the discovery of the secondary flux emergence associated to these eruptions naturally explains the EUV double-loop flaring phenomena. The peculiar J6 was associated to a single magnetic cancellation event, and no other magnetic feature trigger occurred in temporal proximity to the strong  EUV eruption (see Fig. \ref{fig-bzcont-aia}, top) leading us to assume, given the lack of other evidence, that the eruption was impulsive enough to also destabilize the newer neighboring magnetic fields.

For reference, we measure the dipole footpoints' peak strength after the J2 and J3 jet events at 01:36TAI and find strengths of -320$\pm$75 G and 380$\pm$80 G. This newly emerged dipole will subsequently interact with the pore.

Our complete study revealed that six of the studied events (J2, J3, J5, J9, J10, and the uncertain J8) were linked to flux emergence occurrences. 

In the case of the J8 jet, the association with a flux emergence scenario was uncertain because the observations showed very strong magnetic flux emerging at the edge of the PIL, yet rapidly decaying afterwards. In the case of J9 and J10, a smaller magnetic dipole emerged in the very close proximity of an established pore. This explains the more compact nature of the AIA flaring. We assume that this magnetic structure is the cause of both the J9 and J10 events, and probably even the O6 and O7 non-jet flaring events. The individual micro-flare to magnetic field association for these 2 events is unreliable inside our observational (temporal, spatial, and statistical) constraints. In general, we base our determinations on the temporal uniqueness of the emergence events. By that we understand that we identify the only emerging magnetic structure in the selected time frame occurring in the B1 region at or very near ($<2Mm$) the EUV flaring location. We remind the reader that the temporal association is influenced by the data cadence constraints.

\subsection{Minor flaring geyser site events}\label{sec-mmf:sub-other}

\indent\indent Inside our geyser region of interest (B1) we also detected what we defined as minor flaring events (O1-O7, see fig. \ref{fig-bzcont-aia}) recorded in the SDO EUV imager which did not generate detectable jet eruptions. We also recovered photospheric magnetic features linked to these events. In the EUV channels, all O1-O7 events were significantly weaker than any of the J1-J10 jets.  These events, highlighted in Fig. \ref{fig-bzcont-aia} (top) had very weak or no emission in 'hotter' EUV filters (e.g. AIA-94\AA{}), had shorter temporal lifetime and were more constrained spatially. 

A possible explanation of why these events did not trigger detectable EUV jets could revolve around the lack of favorable connections to the open fields of the penumbral canopy structure. The emergent dipole that triggered the J2 and J3 eruptions appears to also be involved in producing some of the smaller flaring events (O1-O2) that did not generate jet emission. This time, we find that small advected MMFs cancel with the new dipole footpoints. The EUV flaring presents a much more localized manifestation, e.g. only the upper part of the structure manifests. 

Short repeated transient flaring characterized the O6 and O7 events. The features were correlated to a dipole emerging at a location, separated from the stronger dipole linked to the J9 and J10 events. We can not establish more than the general temporal agreement between this emergence and the very transient flaring present in the AIA filters. The O3 event could not be associated to any magnetic moving or emerging features. Its magnetic manifestation probably occurred at field strengths lower than  our already relaxed conditions. 

A total of three minor flaring events (O1, O2, and O4) were cautiously associated to cancellations of advected MMFs and three events (O5, O6, and O7) were associated to flux emergence. One event (O3) could not be resolved.  

\begin{table*}[!t]
\centering\scriptsize
\caption{The photospheric magnetic field properties of geyser's 10 recurrent jets are presented below. The jet eruption triggers were attributed to either a flux emergence or a flux cancellation mechanism. Detailed individual correlations along with supporting plots can be found in \citet[Appendix 3]{paraschiv2018}. The J7 and J8 events could not be clearly determined and were labeled `u'. The vertical field ($B_z$) for both MMFs and emerging dipoles was estimated as an average of at least four adjacent pixels with similar strengths. In the case of cancellation events the horizontal movement speeds of MMFs were also calculated. The slit based determinations as shown in fig. \ref{fig-slit} are used to estimate the parameters and associations.
}\label{table-dataset}
\begin{tabular}{ccccccccccc}
  \hline\hline
 &J1&J2&J3&J4&J5&J6&J7&J8&J9&J10\\
 \hline
EUV flaring start [HH:MM TAI] &00:49&01:13 &01:19 &02:11&03:52&13:04&17:06&21:19&22:24&22:55\\
HMI $B_z$ event [HH:MM TAI]&00:48&\multicolumn{2} {c} {01:12-01:24}&02:00-02:12&03:48&13:00&17:00&21:12&\multicolumn{2} {c} {22:24-22:48}\\
 Flux emergence&&x&x&&x&&&u&x&x\\
 Flux cancellation&x&&&x&&x&u&&\\
\multirow{ 2}{*}{ $B_z$ field strength [G]} &\multirow{ 2}{*}{ 361$\pm75$}&\multicolumn{2} {c} {378$\pm$76}&\multirow{ 2}{*}{535$\pm$79}&348$\pm$76&\multirow{ 2}{*}{333$\pm60$}&\multirow{ 2}{*}{161$\pm$58}&645$\pm$69&\multicolumn{2} {c} {408$\pm$53} \\
 &&\multicolumn{2} {c} {-321$\pm$75}&&-291$\pm$39&&&-277$\pm$63&\multicolumn{2} {c} {-339$\pm$56}  \\
 MMF speed [km s$^{-1}$]&0.4&&&0.6&&0.5&0.5&&& \\
 \hline
\end{tabular}
\end{table*}

\section{Discussion}\label{sec-disc}

\indent\indent We report the discovery of 10 successive coronal jets that erupted from a unique EUV structure, rooted at the south-eastern penumbral boundary of AR11302. The observations covered a period of 24 hours. The jets all shared the same spatial origin, with the EUV observations initially suggesting a common trigger mechanism. The jets have comparable coronal morphological features, recur at the same location, and follow the same direction of propagation.

These properties of the repeating jets together with the common EUV site leads to our definition of the `coronal geyser' due to the many similarities with its earth counterpart. Of note, this geyser is not unique: in a complementary study \citep{paraschiv2019} we identified and analyzed six distinct geyser sites of recurrent jets, showing that all are sources of non-thermal emission and Type-III radio bursts.

Our initial goal was to reveal if either flux emergence, or alternatively, flux cancellation can account as a photospheric trigger for our complete set of recurrent jets. Currently, the two main magnetic trigger scenarios discussed in this paper have been each attributed in multiple studies to individual jet eruptions and to short recurrent jet episodes (see discussion in sec. \ref{sec-intro}).  Our MMF correlation described in sec. \ref{sec-mmf} showed that both flux emergence and flux cancellation play a crucial role in producing the 10 recurrent jets studied here. The two possible scenarios are attributed to jets in an almost even ratio (see Table \ref{table-dataset}). Qualitatively, the even ratio also appears to hold when discussing the O1-O7 events (sec. \ref{sec-mmf:sub-other}). No scenario was found to dominate. This result is particularly intriguing as we initially expected one singular process to be the cause of all recurrent jets that erupted from the same reconnection site.

We have shown that a combination of penumbral MMFs and small dipole emergences are involved in producing the recurrent jets. The small MMFs could be tracked for intervals in the order of hours before canceling within the geyser structure (see sec. \ref{sec-mmf:sub-canc} and slit S2 in Fig. \ref{fig-slit}). Alternatively, flux emergence associated jets were seen to manifest concurrently with the EUV eruptions, where the observations showed that all EUV reconnection events occurred inside the 720s integration time of the HMI $B_z$ magnetogram in which the emerging dipoles first manifested (see sec. \ref{sec-mmf:sub-emerg} and slit S1 in Fig. \ref{fig-slit}). We cannot comment on any possible magnetic field dynamics that occur on shorter timescales than are available observationally. The possibility that magnetic manifestations are a cause and not an effect of EUV eruptions remains open for further consideration. See \citet{canfield1996} who hypothesized that approaching photospheric footpoints interact as a consequence of the reconnection happening at higher levels of the solar atmosphere. 

We observe that from an EUV point of view, a unique originating small-scale structure, the geyser,  hosted all reconnection events and subsequent jet eruptions. On the other hand, the involvement of both trigger mechanisms in our recurrent jets leads us to conclude that the events are not in fact homologous though they appear so from the EUV point of view. The photosphere is highly dynamic, hosting what we identified as a possible twisted micro-filament located at the edge of the penumbra, where its footprints are constantly moving (fig. \ref{fig-bzcont-aia}). The EUV data discussed in the companion works showed that our jet eruptions are compatible with micro-filament eruptions as they comprised untwisting strands \citep[e.g.][fig.1.4]{paraschiv2018}. The reconnected plasma followed the pre-existing magnetic fields along the open canopy fields that are all rooted in the penumbral region close to where the geyser structure is also anchored. 

In other EUV observations \citep{guo2013,schmieder2013,hong2013}, the jet eruptions showed a similar morphology, with many of these strands undergoing untwisting motions while propagating along the guiding open field lines as we described in our particular case.  The untwisting motions have been associated in the standard jet eruption scenario with newly emerging flux that is released by an interchange reconnection with an open field structure as described by \citet{pariat2009}.  The analysis of \citet{sterling2015} suggest that the standard jet eruption interpretation \citep{moore2010,moore2013} may be unreliable. The standard interpretation assumes ideal conditions and implies that jets can occur in an environment that does not host twisted filaments. On the contrary, the observational evidence links the recurrent EUV jets to the blowout hypothesis involving micro-filament eruptions. Additionally, the micro-filament eruption scenario requires that multiple short-lived ($<$30 s) and fast reconnection events occur in order to constitute an observed eruptive blowout jet as described in \citet{moore2010}. Fig. \ref{fig-bzcont-aia} shows that the EUV timeseries profiles are compatible with the blowout interpretation. This aspect was extensively discussed in the complementary \citet{paraschiv2019} study.

We ask the following: Can a set of recurrent eruptions that are generated via a combination penumbral MMFs and small dipole emergences be compatible with recurring micro-filament eruptions? The observational interpretation of a micro-filament eruption translates to a `store and release' process \citep[e.g. as discussed by][]{cheung2014} involving multiple stages: i. energy is deposited in the site; ii. the site is destabilized; iii. the stored energy is released, usually in the form of eruptions. In our particular case, we conjecture that the photospheric MMFs act as the destabilizing factor for at least jets that resulted from flux cancellation, and that `twist' is transported primarily to the site (particularly in the PIL region) via a micro-filament, though a contribution from an eruption configuration as described by \citet{canfield1996} can not be discarded. This schema is compatible with the erupting mechanism proposed by \citet{sterling2015} and \citet{sterling2017}. 


In the absence of MMF destabilization we note that a series of recurrent jets could be also attributed to a micro-filament that is subjected to energy storage and release. Accurate horizontal fields and current estimations could not be retrieved due to data constraints and uncertainties, limiting us in adding further input. 

Our future work will focus on recovering high quality magnetic and electric fields information at multiple atmospheric heights which will help us  answer this question. Low noise and fast cadence spectro-polarimetric magnetic field measurements of a geyser structure in both the corona and chromosphere will be able to confirm or invalidate our current findings. The commencement of DKIST \citep{tritschler2016} operations will prove invaluable to this goal.

\section{Conclusion}\label{sec-concl}

The interpretation of the multiple magnetic moving features and emergent magnetic fields provides insights into the photospheric manifestations that were involved in generating a series of recurrent jet eruptions associated to a distinct geyser structure. We produced magnetic field data maps which are suitable for the interpretation of small and intermediate flux fields that enabled us to recover stable patches of small-scale photospheric vertical magnetic fields.  

We attempted to identify a single process linked to the formation of recurrent jets that originateed from the same reconnection site, the geyser. We showed that at penumbral small-scales both widely discussed magnetic trigger mechanisms, namely flux emergence and flux cancellation, were involved in generating recurrent jets over an impressively long observation period of 24 hours. To our knowledge, this is the first study to support that generating recurrent EUV/X-Ray jets is possible via a combination of two distinct of photospheric magnetic processes. Neither of the two had been found to dominate. Additionally, we found evidence that a scenario involving micro-filament eruptions is also compatible with the observational evidence. 

By assuming that solar reconnection processes are scale independent, this analysis of small-scale eruptions provides insight towards the understanding of the more widely debated large scale eruptive events.

\acknowledgments
A.R.P. acknowledges support through Monash University, The Monash School of Mathematical Sciences, the Astronomical Society of Australia and through an Australian Government Research Training Program (RTP) Scholarship. A.R.P. also recognizes the very productive discussions and training that took place at NWRA and would like to thank Dr. Graham Barnes and Dr. Charles Lindsey for the support offered. Raw data and calibration instructions are obtained courtesy of NASA/SDO-HMI, SDO-AIA, and STEREO-EUVI  science teams. The authors welcome and appreciate the open data policy of the SDO and STEREO missions. The authors thank the anonymous reviewers for their input.

\bibliography{bibliography}

\begin{thebibliography}{}
\expandafter\ifx\csname natexlab\endcsname\relax\def\natexlab#1{#1}\fi
\providecommand{\url}[1]{\href{#1}{#1}}
\providecommand{\dodoi}[1]{doi:~\href{http://doi.org/#1}{\nolinkurl{#1}}}
\providecommand{\doeprint}[1]{\href{http://ascl.net/#1}{\nolinkurl{http://ascl.net/#1}}}
\providecommand{\doarXiv}[1]{\href{https://arxiv.org/abs/#1}{\nolinkurl{https://arxiv.org/abs/#1}}}

\bibitem[{{Alissandrakis}(1981)}]{alissandrakis1981}
{Alissandrakis}, C.~E. 1981, \aap, 100, 197

\bibitem[{{Archontis} \& {Hood}(2013)}]{archontis2013}
{Archontis}, V., \& {Hood}, A.~W. 2013, \apjl, 769, L21,
  \dodoi{10.1088/2041-8205/769/2/L21}

\bibitem[{{Archontis} {et~al.}(2010){Archontis}, {Tsinganos}, \&
  {Gontikakis}}]{archontis2010}
{Archontis}, V., {Tsinganos}, K., \& {Gontikakis}, C. 2010, \aap, 512, L2,
  \dodoi{10.1051/0004-6361/200913752}

\bibitem[{Bobra {et~al.}(2014)Bobra, Sun, Hoeksema, Turmon, Liu, Hayashi,
  Barnes, \& Leka}]{bobra2014}
Bobra, M.~G., Sun, X., Hoeksema, J.~T., {et~al.} 2014, Solar Physics, 289,
  3549, \dodoi{10.1007/s11207-014-0529-3}

\bibitem[{{Canfield} {et~al.}(1996){Canfield}, {Reardon}, {Leka}, {Shibata},
  {Yokoyama}, \& {Shimojo}}]{canfield1996}
{Canfield}, R.~C., {Reardon}, K.~P., {Leka}, K.~D., {et~al.} 1996, \apj, 464,
  1016, \dodoi{10.1086/177389}

\bibitem[{{Chandra} {et~al.}(2015){Chandra}, {Gupta}, {Mulay}, \&
  {Tripathi}}]{Chandra2015}
{Chandra}, R., {Gupta}, G.~R., {Mulay}, S., \& {Tripathi}, D. 2015, \mnras,
  446, 3741, \dodoi{10.1093/mnras/stu2305}

\bibitem[{{Chen} {et~al.}(2015{\natexlab{a}}){Chen}, {Zhang}, {Ma}, {Yang},
  {Li}, {Huang}, \& {Xiao}}]{chenh2015}
{Chen}, H., {Zhang}, J., {Ma}, S., {et~al.} 2015{\natexlab{a}}, \apjl, 808,
  L24, \dodoi{10.1088/2041-8205/808/1/L24}

\bibitem[{{Chen} {et~al.}(2008){Chen}, {Jiang}, \& {Ma}}]{chen2008}
{Chen}, H.~D., {Jiang}, Y.~C., \& {Ma}, S.~L. 2008, \aap, 478, 907,
  \dodoi{10.1051/0004-6361:20078641}

\bibitem[{{Chen} {et~al.}(2015{\natexlab{b}}){Chen}, {Su}, {Yin}, {Priya},
  {Zhang}, {Liu}, {Xu}, \& {Yu}}]{chen2015}
{Chen}, J., {Su}, J., {Yin}, Z., {et~al.} 2015{\natexlab{b}}, \apj, 815, 71,
  \dodoi{10.1088/0004-637X/815/1/71}

\bibitem[{{Cheung} \& {Isobe}(2014)}]{cheung2014}
{Cheung}, M.~C.~M., \& {Isobe}, H. 2014, Living Reviews in Solar Physics, 11,
  3, \dodoi{10.12942/lrsp-2014-3}

\bibitem[{{Cheung} {et~al.}(2015){Cheung}, {De Pontieu}, {Tarbell}, {Fu},
  {Tian}, {Testa}, {Reeves}, {Mart{\'{\i}}nez-Sykora}, {Boerner}, {W{\"u}lser},
  {Lemen}, {Title}, {Hurlburt}, {Kleint}, {Kankelborg}, {Jaeggli}, {Golub},
  {McKillop}, {Saar}, {Carlsson}, \& {Hansteen}}]{cheung2015}
{Cheung}, M.~C.~M., {De Pontieu}, B., {Tarbell}, T.~D., {et~al.} 2015, \apj,
  801, 83, \dodoi{10.1088/0004-637X/801/2/83}

\bibitem[{{Falconer} {et~al.}(2016){Falconer}, {Tiwari}, {Moore}, \&
  {Khazanov}}]{falconer2016}
{Falconer}, D.~A., {Tiwari}, S.~K., {Moore}, R.~L., \& {Khazanov}, I. 2016,
  \apjl, 833, L31, \dodoi{10.3847/2041-8213/833/2/L31}

\bibitem[{{Freeland} \& {Handy}(1998)}]{freeland1998}
{Freeland}, S.~L., \& {Handy}, B.~N. 1998, \solphys, 182, 497,
  \dodoi{10.1023/A:1005038224881}

\bibitem[{{Gary} \& {Hagyard}(1990)}]{gary1990}
{Gary}, G.~A., \& {Hagyard}, M.~J. 1990, \solphys, 126, 21,
  \dodoi{10.1007/BF00158295}

\bibitem[{{Guo} {et~al.}(2013){Guo}, {D{\'e}moulin}, {Schmieder}, {Ding},
  {Vargas Dom{\'{\i}}nguez}, \& {Liu}}]{guo2013}
{Guo}, Y., {D{\'e}moulin}, P., {Schmieder}, B., {et~al.} 2013, \aap, 555, A19,
  \dodoi{10.1051/0004-6361/201321229}

\bibitem[{{Hanson} {et~al.}(2015){Hanson}, {Donea}, \& {Leka}}]{hanson2015}
{Hanson}, C.~S., {Donea}, A.~C., \& {Leka}, K.~D. 2015, \solphys, 290, 2171,
  \dodoi{10.1007/s11207-015-0743-7}

\bibitem[{{Harvey}(1969)}]{harvey1969}
{Harvey}, J.~W. 1969, PhD thesis, National Solar Observatory

\bibitem[{{Harvey} \& {Harvey}(1973)}]{harvey1973}
{Harvey}, K., \& {Harvey}, J. 1973, \solphys, 28, 61,
  \dodoi{10.1007/BF00152912}

\bibitem[{{Hoeksema} {et~al.}(2014){Hoeksema}, {Liu}, {Hayashi}, {Sun},
  {Schou}, {Couvidat}, {Norton}, {Bobra}, {Centeno}, {Leka}, {Barnes}, \&
  {Turmon}}]{hoeksema2014}
{Hoeksema}, J.~T., {Liu}, Y., {Hayashi}, K., {et~al.} 2014, \solphys, 289,
  3483, \dodoi{10.1007/s11207-014-0516-8}

\bibitem[{{Hong} {et~al.}(2013){Hong}, {Jiang}, {Yang}, {Zheng}, {Bi}, {Li},
  {Yang}, \& {Yang}}]{hong2013}
{Hong}, J.-C., {Jiang}, Y.-C., {Yang}, J.-Y., {et~al.} 2013, Research in
  Astronomy and Astrophysics, 13, 253, \dodoi{10.1088/1674-4527/13/3/001}

\bibitem[{{Jiang} {et~al.}(2007){Jiang}, {Chen}, {Li}, {Shen}, \&
  {Yang}}]{jiang2007}
{Jiang}, Y.~C., {Chen}, H.~D., {Li}, K.~J., {Shen}, Y.~D., \& {Yang}, L.~H.
  2007, \aap, 469, 331, \dodoi{10.1051/0004-6361:20053954}

\bibitem[{{Kaiser} {et~al.}(2008){Kaiser}, {Kucera}, {Davila}, {St.~Cyr},
  {Guhathakurta}, \& {Christian}}]{kaiser2008}
{Kaiser}, M.~L., {Kucera}, T.~A., {Davila}, J.~M., {et~al.} 2008, \ssr, 136, 5,
  \dodoi{10.1007/s11214-007-9277-0}

\bibitem[{{Kosugi} {et~al.}(2007){Kosugi}, {Matsuzaki}, {Sakao}, {Shimizu},
  {Sone}, {Tachikawa}, {Hashimoto}, {Minesugi}, {Ohnishi}, {Yamada}, {Tsuneta},
  {Hara}, {Ichimoto}, {Suematsu}, {Shimojo}, {Watanabe}, {Shimada}, {Davis},
  {Hill}, {Owens}, {Title}, {Culhane}, {Harra}, {Doschek}, \&
  {Golub}}]{kosugi2007}
{Kosugi}, T., {Matsuzaki}, K., {Sakao}, T., {et~al.} 2007, \solphys, 243, 3,
  \dodoi{10.1007/s11207-007-9014-6}

\bibitem[{{Lee} {et~al.}(2015){Lee}, {Archontis}, \& {Hood}}]{lee2015}
{Lee}, E.~J., {Archontis}, V., \& {Hood}, A.~W. 2015, \apjl, 798, L10,
  \dodoi{10.1088/2041-8205/798/1/L10}

\bibitem[{{Leka} {et~al.}(2009){Leka}, {Barnes}, {Crouch}, {Metcalf}, {Gary},
  {Jing}, \& {Liu}}]{leka2009}
{Leka}, K.~D., {Barnes}, G., {Crouch}, A.~D., {et~al.} 2009, \solphys, 260, 83,
  \dodoi{10.1007/s11207-009-9440-8}

\bibitem[{{Leka} {et~al.}(1994){Leka}, {Canfield}, {Mickey}, {van
  Driel-Gesztelyi}, {Nitta}, {Sakurai}, \& {Ichimoto}}]{leka1994}
{Leka}, K.~D., {Canfield}, R.~C., {Mickey}, D.~L., {et~al.} 1994, \solphys,
  155, 301, \dodoi{10.1007/BF00680598}

\bibitem[{{Lemen} {et~al.}(2012){Lemen}, {Title}, {Akin}, {Boerner}, {Chou},
  {Drake}, {Duncan}, {Edwards}, {Friedlaender}, {Heyman}, {Hurlburt}, {Katz},
  {Kushner}, {Levay}, {Lindgren}, {Mathur}, {McFeaters}, {Mitchell}, {Rehse},
  {Schrijver}, {Springer}, {Stern}, {Tarbell}, {Wuelser}, {Wolfson}, {Yanari},
  {Bookbinder}, {Cheimets}, {Caldwell}, {Deluca}, {Gates}, {Golub}, {Park},
  {Podgorski}, {Bush}, {Scherrer}, {Gummin}, {Smith}, {Auker}, {Jerram},
  {Pool}, {Soufli}, {Windt}, {Beardsley}, {Clapp}, {Lang}, \&
  {Waltham}}]{lemen2012}
{Lemen}, J.~R., {Title}, A.~M., {Akin}, D.~J., {et~al.} 2012, \solphys, 275,
  17, \dodoi{10.1007/s11207-011-9776-8}

\bibitem[{{Li} {et~al.}(2015){Li}, {Yang}, {Chen}, {Li}, \& {Zhang}}]{li2015}
{Li}, X., {Yang}, S., {Chen}, H., {Li}, T., \& {Zhang}, J. 2015, \apjl, 814,
  L13, \dodoi{10.1088/2041-8205/814/1/L13}

\bibitem[{{Martin} {et~al.}(1985){Martin}, {Livi}, \& {Wang}}]{martin1985}
{Martin}, S.~F., {Livi}, S.~H.~B., \& {Wang}, J. 1985, Australian Journal of
  Physics, 38, 929, \dodoi{10.1071/PH850929}

\bibitem[{{Metcalf}(1994)}]{metcalf1994}
{Metcalf}, T.~R. 1994, \solphys, 155, 235, \dodoi{10.1007/BF00680593}

\bibitem[{{Metcalf} {et~al.}(2006){Metcalf}, {Leka}, {Barnes}, {Lites},
  {Georgoulis}, {Pevtsov}, {Balasubramaniam}, {Gary}, {Jing}, {Li}, {Liu},
  {Wang}, {Abramenko}, {Yurchyshyn}, \& {Moon}}]{metcalf2006}
{Metcalf}, T.~R., {Leka}, K.~D., {Barnes}, G., {et~al.} 2006, \solphys, 237,
  267, \dodoi{10.1007/s11207-006-0170-x}

\bibitem[{{Moore} {et~al.}(2010){Moore}, {Cirtain}, {Sterling}, \&
  {Falconer}}]{moore2010}
{Moore}, R.~L., {Cirtain}, J.~W., {Sterling}, A.~C., \& {Falconer}, D.~A. 2010,
  \apj, 720, 757, \dodoi{10.1088/0004-637X/720/1/757}

\bibitem[{{Moore} {et~al.}(2013){Moore}, {Sterling}, {Falconer}, \&
  {Robe}}]{moore2013}
{Moore}, R.~L., {Sterling}, A.~C., {Falconer}, D.~A., \& {Robe}, D. 2013, \apj,
  769, 134, \dodoi{10.1088/0004-637X/769/2/134}

\bibitem[{{Moreno-Insertis} \& {Galsgaard}(2013)}]{moreno2013}
{Moreno-Insertis}, F., \& {Galsgaard}, K. 2013, \apj, 771, 20,
  \dodoi{10.1088/0004-637X/771/1/20}

\bibitem[{{Moreno-Insertis} {et~al.}(2008){Moreno-Insertis}, {Galsgaard}, \&
  {Ugarte-Urra}}]{moreno2008}
{Moreno-Insertis}, F., {Galsgaard}, K., \& {Ugarte-Urra}, I. 2008, \apjl, 673,
  L211, \dodoi{10.1086/527560}

\bibitem[{{Panesar} {et~al.}(2016){Panesar}, {Sterling}, \&
  {Moore}}]{panesar2016}
{Panesar}, N.~K., {Sterling}, A.~C., \& {Moore}, R.~L. 2016, \apjl, 822, L23,
  \dodoi{10.3847/2041-8205/822/2/L23}

\bibitem[{Paraschiv {et~al.}(2019)Paraschiv, Donea, \&
  Leka}]{paraschiv2019-data}
Paraschiv, A., Donea, A., \& Leka, K. 2019, {Replication data and minimal code
  examples for: ``The Trigger Mechanism of Recurrent Active Region Jets
  Revealed by the Magnetic Properties of a Coronal Geyser Site''}, V1,  Harvard
  Dataverse, \dodoi{10.7910/DVN/USRJXX}.
\newblock \url{https://doi.org/10.7910/DVN/USRJXX}

\bibitem[{Paraschiv(2018)}]{paraschiv2018}
Paraschiv, A.~R. 2018, \dodoi{10.26180/5bc9d76627396}

\bibitem[{Paraschiv \& Donea(2019)}]{paraschiv2019}
Paraschiv, A.~R., \& Donea, A. 2019, The Astrophysical Journal, 873, 110,
  \dodoi{10.3847/1538-4357/ab04a6}

\bibitem[{{Pariat} {et~al.}(2009){Pariat}, {Antiochos}, \&
  {DeVore}}]{pariat2009}
{Pariat}, E., {Antiochos}, S.~K., \& {DeVore}, C.~R. 2009, \apj, 691, 61,
  \dodoi{10.1088/0004-637X/691/1/61}

\bibitem[{{Pariat} {et~al.}(2010){Pariat}, {Antiochos}, \&
  {DeVore}}]{pariat2010}
---. 2010, \apj, 714, 1762, \dodoi{10.1088/0004-637X/714/2/1762}

\bibitem[{{Pariat} {et~al.}(2015){Pariat}, {Dalmasse}, {DeVore}, {Antiochos},
  \& {Karpen}}]{pariat2015}
{Pariat}, E., {Dalmasse}, K., {DeVore}, C.~R., {Antiochos}, S.~K., \& {Karpen},
  J.~T. 2015, \aap, 573, A130, \dodoi{10.1051/0004-6361/201424209}

\bibitem[{{Pariat} {et~al.}(2016){Pariat}, {Dalmasse}, {DeVore}, {Antiochos},
  \& {Karpen}}]{pariat2016}
---. 2016, \aap, 596, A36, \dodoi{10.1051/0004-6361/201629109}

\bibitem[{{Pesnell} {et~al.}(2012){Pesnell}, {Thompson}, \&
  {Chamberlin}}]{pesnell2012}
{Pesnell}, W.~D., {Thompson}, B.~J., \& {Chamberlin}, P.~C. 2012, \solphys,
  275, 3, \dodoi{10.1007/s11207-011-9841-3}

\bibitem[{{Raouafi} {et~al.}(2016){Raouafi}, {Patsourakos}, {Pariat}, {Young},
  {Sterling}, {Savcheva}, {Shimojo}, {Moreno-Insertis}, {DeVore}, {Archontis},
  {T{\"o}r{\"o}k}, {Mason}, {Curdt}, {Meyer}, {Dalmasse}, \&
  {Matsui}}]{raouafi2016}
{Raouafi}, N.~E., {Patsourakos}, S., {Pariat}, E., {et~al.} 2016, \ssr, 201, 1,
  \dodoi{10.1007/s11214-016-0260-5}

\bibitem[{{Sakaue} {et~al.}(2017){Sakaue}, {Tei}, {Asai}, {Ueno}, {Ichimoto},
  \& {Shibata}}]{sakaue2017}
{Sakaue}, T., {Tei}, A., {Asai}, A., {et~al.} 2017, \pasj, 69, 80,
  \dodoi{10.1093/pasj/psx071}

\bibitem[{{Sakaue} {et~al.}(2018){Sakaue}, {Tei}, {Asai}, {Ueno}, {Ichimoto},
  \& {Shibata}}]{sakaue2018}
---. 2018, \pasj, \dodoi{10.1093/pasj/psx133}

\bibitem[{{Scherrer} {et~al.}(2012){Scherrer}, {Schou}, {Bush}, {Kosovichev},
  {Bogart}, {Hoeksema}, {Liu}, {Duvall}, {Zhao}, {Title}, {Schrijver},
  {Tarbell}, \& {Tomczyk}}]{scherrer2012}
{Scherrer}, P.~H., {Schou}, J., {Bush}, R.~I., {et~al.} 2012, \solphys, 275,
  207, \dodoi{10.1007/s11207-011-9834-2}

\bibitem[{{Schmieder} {et~al.}(2013){Schmieder}, {Guo}, {Moreno-Insertis},
  {Aulanier}, {Yelles Chaouche}, {Nishizuka}, {Harra}, {Thalmann}, {Vargas
  Dominguez}, \& {Liu}}]{schmieder2013}
{Schmieder}, B., {Guo}, Y., {Moreno-Insertis}, F., {et~al.} 2013, \aap, 559,
  A1, \dodoi{10.1051/0004-6361/201322181}

\bibitem[{{Shimojo} {et~al.}(1996){Shimojo}, {Hashimoto}, {Shibata},
  {Hirayama}, {Hudson}, \& {Acton}}]{shimojo1996}
{Shimojo}, M., {Hashimoto}, S., {Shibata}, K., {et~al.} 1996, \pasj, 48, 123

\bibitem[{{Shimojo} \& {Shibata}(2000)}]{shimojo2000}
{Shimojo}, M., \& {Shibata}, K. 2000, \apj, 542, 1100, \dodoi{10.1086/317024}

\bibitem[{{St.~Cyr} {et~al.}(1997){St.~Cyr}, {Howard}, {Simnett}, {Gurman},
  {Plunkett}, {Sheeley}, {Schwenn}, {Koomen}, {Brueckner}, {Michels},
  {Andrews}, {Biesecker}, {Cook}, {Dere}, {Duffin}, {Einfalt}, {Korendyke},
  {Lamy}, {Lewis}, {Llebaria}, {Lyons}, {Moses}, {Moulton}, {Newmark},
  {Paswaters}, {Podlipnik}, {Rich}, {Schenk}, {Socker}, {Stezelberger},
  {Tappin}, {Thompson}, \& {Wang}}]{stcyr1997}
{St.~Cyr}, O.~C., {Howard}, R.~A., {Simnett}, G.~M., {et~al.} 1997, in ESA
  Special Publication, Vol. 415, Correlated Phenomena at the Sun, in the
  Heliosphere and in Geospace, ed. A.~{Wilson}, 103

\bibitem[{{Sterling} {et~al.}(2015){Sterling}, {Moore}, {Falconer}, \&
  {Adams}}]{sterling2015}
{Sterling}, A.~C., {Moore}, R.~L., {Falconer}, D.~A., \& {Adams}, M. 2015,
  \nat, 523, 437, \dodoi{10.1038/nature14556}

\bibitem[{{Sterling} {et~al.}(2016){Sterling}, {Moore}, {Falconer}, {Panesar},
  {Akiyama}, {Yashiro}, \& {Gopalswamy}}]{sterling2016}
{Sterling}, A.~C., {Moore}, R.~L., {Falconer}, D.~A., {et~al.} 2016, \apj, 821,
  100, \dodoi{10.3847/0004-637X/821/2/100}

\bibitem[{{Sterling} {et~al.}(2017){Sterling}, {Moore}, {Falconer}, {Panesar},
  \& {Martinez}}]{sterling2017}
{Sterling}, A.~C., {Moore}, R.~L., {Falconer}, D.~A., {Panesar}, N.~K., \&
  {Martinez}, F. 2017, \apj, 844, 28, \dodoi{10.3847/1538-4357/aa7945}

\bibitem[{{Tritschler} {et~al.}(2016){Tritschler}, {Rimmele}, {Berukoff},
  {Casini}, {Kuhn}, {Lin}, {Rast}, {McMullin}, {Schmidt}, {W{\"o}ger}, \&
  {DKIST Team}}]{tritschler2016}
{Tritschler}, A., {Rimmele}, T.~R., {Berukoff}, S., {et~al.} 2016,
  Astronomische Nachrichten, 337, 1064, \dodoi{10.1002/asna.201612434}

\bibitem[{{Tsuneta} {et~al.}(2008){Tsuneta}, {Ichimoto}, {Katsukawa}, {Nagata},
  {Otsubo}, {Shimizu}, {Suematsu}, {Nakagiri}, {Noguchi}, {Tarbell}, {Title},
  {Shine}, {Rosenberg}, {Hoffmann}, {Jurcevich}, {Kushner}, {Levay}, {Lites},
  {Elmore}, {Matsushita}, {Kawaguchi}, {Saito}, {Mikami}, {Hill}, \&
  {Owens}}]{tsuneta2008}
{Tsuneta}, S., {Ichimoto}, K., {Katsukawa}, Y., {et~al.} 2008, \solphys, 249,
  167, \dodoi{10.1007/s11207-008-9174-z}

\bibitem[{{van Driel-Gesztelyi} \& {Green}(2015)}]{vandrielgesztelyi2015}
{van Driel-Gesztelyi}, L., \& {Green}, L.~M. 2015, Living Reviews in Solar
  Physics, 12, 1, \dodoi{10.1007/lrsp-2015-1}

\bibitem[{{Wang} {et~al.}(1998){Wang}, {Sheeley}, {Socker}, {Howard},
  {Brueckner}, {Michels}, {Moses}, {St.~Cyr}, {Llebaria}, \&
  {Delaboudini{\`e}re}}]{wang1998}
{Wang}, Y.-M., {Sheeley}, Jr., N.~R., {Socker}, D.~G., {et~al.} 1998, \apj,
  508, 899, \dodoi{10.1086/306450}

\bibitem[{{Wuelser} {et~al.}(2004){Wuelser}, {Lemen}, {Tarbell}, {Wolfson},
  {Cannon}, {Carpenter}, {Duncan}, {Gradwohl}, {Meyer}, {Moore}, {Navarro},
  {Pearson}, {Rossi}, {Springer}, {Howard}, {Moses}, {Newmark},
  {Delaboudiniere}, {Artzner}, {Auchere}, {Bougnet}, {Bouyries}, {Bridou},
  {Clotaire}, {Colas}, {Delmotte}, {Jerome}, {Lamare}, {Mercier}, {Mullot},
  {Ravet}, {Song}, {Bothmer}, \& {Deutsch}}]{wuelser2004}
{Wuelser}, J.-P., {Lemen}, J.~R., {Tarbell}, T.~D., {et~al.} 2004, in Society
  of Photo-Optical Instrumentation Engineers (SPIE) Conference Series, Vol.
  5171, Telescopes and Instrumentation for Solar Astrophysics, ed.
  S.~{Fineschi} \& M.~A. {Gummin}, 111--122

\bibitem[{{Wyper} {et~al.}(2018){Wyper}, {DeVore}, \& {Antiochos}}]{wyper2018}
{Wyper}, P.~F., {DeVore}, C.~R., \& {Antiochos}, S.~K. 2018, \apj, 852, 98,
  \dodoi{10.3847/1538-4357/aa9ffc}

\end{thebibliography}



\end{document}